\documentstyle[12pt,epsfig]{article}

\def\0{\over } \def\1{\vec }     \def\2{{1\over2}} \def\4{{1\over4}}            
\def\5{\bar }  \def\6{\partial } \def\7#1{{#1}\llap{/}}                         
\def\8#1{{\textstyle{#1}}}       \def\9#1{{\bf {#1}}}                           
 \def\llp{\hbox to 0pt{\hss /\hskip1.5pt}}    
\def\llo{\hbox to 0.2pt{\hss /}} \def\llq{\hbox to 0pt{\hss /\hskip0.5pt}}      
\def\so{\supset\hbox to 0pt{\hss $\displaystyle -$\hskip1pt}}

\def\<{\langle } \def\>{\rangle }

\let\nn=\nonumber                                                               
\def\bea{\begin{eqnarray}} \def\eea{\end{eqnarray}}                             
\def\beann{\begin{eqnarray*}} \def\eeann{\end{eqnarray*}}                       
\def\beq{\begin{equation}} \def\eeq{\end{equation}}

\newcommand{\FF}{\mbox{$F_2$}}
\newcommand{\xB}{\mbox{$x~$}}    
\newcommand{\xinv}{\mbox{$\frac{1}{x}$}}    
\newcommand{\QQ}{\mbox{$Q^2$}}
\newcommand{\QF}{\mbox{$Q_0^2$}}
\newcommand{\xf}{\mbox{$x_0$}}
\newcommand{\unit}[1]{{\rm \,#1}}
\newcommand{\GeV}{\unit{GeV}}

\date{}
\title{
{\large\rm DESY 96-061}\hfill{\large\tt ISSN 0418-9833}\\
{\large\rm May 1996}\hfill\vspace*{3.5cm}\\
Double-logarithmic Scaling\\
of the Structure Function $F_2$ at small $x$}
\author{W. Buchm\"{u}ller and D. Haidt\\
{\normalsize\it Deutsches Elektronen-Synchrotron DESY, 22603 Hamburg, Germany}
\vspace*{3.5cm}\\                     
}                                                                             
\begin{document}                                                                

\setlength{\baselineskip}{18pt}                                     
\maketitle  
\thispagestyle{empty}
\begin{abstract}
\noindent
Recent data on the structure function $F_2(x,Q^2)$ at small values of $x$
are analysed and compared with theoretical expectations. It is shown that
the observed rise at small $x$ is consistent with a logarithmic
increase, growing 
logarithmically also with $Q^2$. A stronger increase, which may
be incompatible with unitarity when extrapolated to asymptotically small
values of $x$, cannot be inferred from present data.
\end{abstract} 
\newpage                                             
\section{Introduction}
%
The H1 and ZEUS collaborations at the $ep$ collider
HERA have published measurements of the structure function $F_2(x,Q^2)$
\cite{h11,zeus,h12}, which extend the range in $x$ and $Q^2$ by two orders of
magnitude as compared to previous fixed target experiments. 
A prominent rise of the structure function has been observed at small
values of $x$, which has stimulated a variety of theoretical investigations.

The behaviour of the structure function $F_2(x,Q^2)$ for 
$x\rightarrow 0$, with $Q^2$ fixed, corresponds
to the high-energy or Regge limit for virtual photon-proton scattering.
The computation of the high-energy behaviour of cross sections in QCD is an
important and still unsolved problem. Theoretical approaches
to understand the Regge limit are almost exclusively based on perturbation
theory. The standard evolution equations\footnote{For a review, see 
\cite{ynd}.} predict a rise of the structure function $F_2$ at small $x$.
This holds for the double-asymptotic solution of the evolution equations
at large $Q^2$ and small $x$ \cite{der}, as well as for the solution of
the BFKL equation \cite{lip} which makes a prediction for the growth
at small $x$ for fixed $Q^2$. One expects that the rise of cross sections
at large energies is eventually damped by screening corrections \cite{levin} 
leading to an asymptotic behaviour which, for proton-proton scattering,
has to satisfy the Froissart bound \cite{fross}.

The starting point of this paper is the examination of the small-\xB\ regime
of the HERA data. For values of \xB\ below about $10^{-2}$ all measurements are 
compatible with a double-logarithmic behaviour in \xB\ and \QQ. 
We then address the question
whether the observed rise of the structure function $F_2$ at 
small $x$ is consistent with unitarity bounds. It is argued 
that a $\ln{1\over x}$ increase, if persistent to asymptotically small \xB,
may be compatible with constraints from unitarity. The double-logarithmic
fit to the data is compared with the double-asymptotic form considered by
Ball and Forte \cite{bf1,bf2}.
%
\section{Phenomenological analysis}
\begin{figure}[hbtp]\centering
   \mbox{\epsfig{file=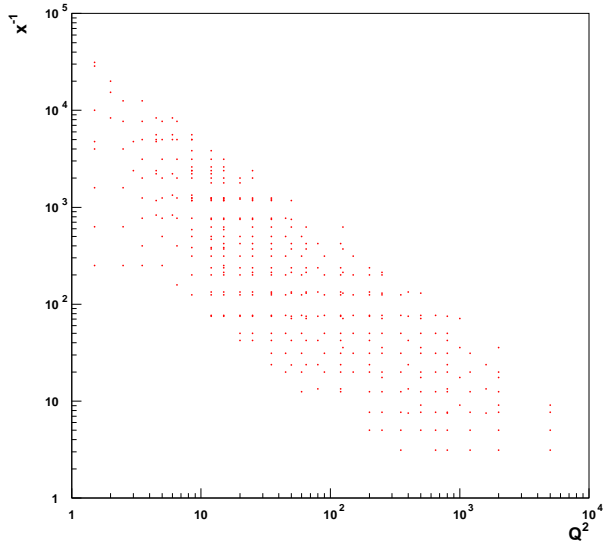,width=8.6cm}}
   \caption{\sl (\QQ,\xinv)-phasespace region covered by the H1 and ZEUS 
            measurements of \FF.}
   \label{fig:phsp}
\end{figure}

%
The collaborations H1 and ZEUS have provided measurements at various values of 
\QQ\ covering the range in \xB\ as shown in fig.~\ref{fig:phsp}. 
The measurements of both collaborations are compatible with each other.
The 1994 data by the H1 collaboration \cite{h12} are used for the fits,
since at present they are the most precise ones, and they also 
allow the distinction
between correlated and uncorrelated contributions to the systematic
uncertainties.

\begin{figure}[hbtp]\centering
   \mbox{\epsfig{file=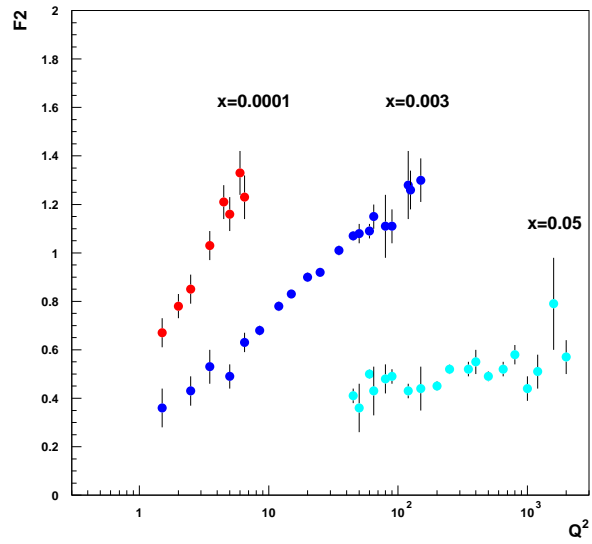,width=8.6cm}}
   \caption{\sl $F(\xB,\QQ)$ versus $Q^2$ for three values of \xB : upper
            points 0.0001, middle points 0.003, lower points 0.05. }
   \label{fig:xfix}
\end{figure}

\begin{figure}[hbtp]\centering
   \mbox{\epsfig{file=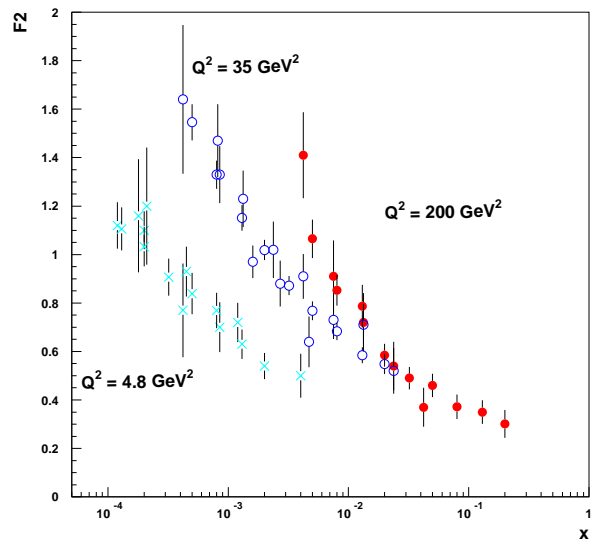,width=8.6cm}}
   \caption{\sl $F(\xB,\QQ)$ versus $x$ for three values of $Q^2$ : 4.8,
            35 and 200 GeV$^2$. }
   \label{fig:q2fix}
\end{figure}

The behaviour of the structure function \FF\ in the covered kinematical
range is displayed in figs.~\ref{fig:xfix} and \ref{fig:q2fix}  
as function of \xB\ and \QQ, respectively. 
The \QQ-dependence for three bands in $x$, displayed 
in fig.~\ref{fig:xfix}, shows a logarithmic behaviour. The slopes 
are strongly \xB-dependent. For  $\xB \simeq$ 0.05 the slope vanishes 
and the structure function \FF\ becomes independent of \QQ. 
Fig.~\ref{fig:q2fix} shows the \xB-behaviour of \FF\ for three values of
\QQ. The striking new feature of the HERA data is the prominent rise
at values of \xB\ below about 10$^{-2}$. This small-\xB\ regime 
connects to the valence region, which was intensively investigated in 
previous low energy experiments. 
 
In the following only the small-\xB\ data will be considered. The restricted 
phase space region is defined by 
\beq
 \xB\ < 0.010\quad,\quad \QQ\ > 5\ \mbox{GeV}^2\qquad. \label{eq:acc}
\eeq
The cut in \xB\ implies an effective cut of the high \QQ-data 
(cf. fig.~\ref{fig:phsp}). 

It is a remarkable property of the data that in this small-\xB\ regime
they are well described by a 
polynomial linear in $\log{\xinv}$ for every measured \QQ-value, 
\begin{equation}
      \FF(x,Q^2) = u_0(Q^2) +u_1(Q^2)\ \log{\frac{v(\QQ)}{x}} \ .
      \label{eq:logx}
\end{equation}
Possible higher order terms in $\log{\xinv}$ are statistically not signi\-ficant
and are consequently not considered. The quantities $u_0$ and $u_1$ are 
functions of \QQ, as well as the quantity $\log{v}$, which is uniquely 
defined as the weighted average $\langle\log{\xinv}\rangle$. 
Using uncorrelated errors,
also the uncertainties of $u_0$ and $u_1$ are uncorrelated. The numerical 
value of $v(\QQ)$ reflects the available $x$-range for any given $Q^2$ 
(see fig.~\ref{fig:phsp}), as well as the precision of the data. 
It is then possible 
to represent the whole body of data in the restricted phase space region, 
defined by (\ref{eq:acc}), for each measured \QQ-value by three numbers,
\beq
v(\QQ)\ ,\quad u_0(\QQ) \pm \delta u_0(\QQ)\ ,\quad
u_1(\QQ) \pm \delta u_1(\QQ)\ .
\label{eq:data} 
\eeq
In terms of \FF\ the two independent functions $u_0$ and $u_1$ represent
average and slope,
\begin{eqnarray}
 u_0(\QQ) &=& \FF(v(\QQ),\QQ)\ ,    \nonumber \\
 u_1(\QQ) &=& \frac{\partial}{\partial\log{\xinv}}\FF(\xB,\QQ)\ . 
 \label{eq:ui} 
\end{eqnarray} 

The data on \FF\ is consistent with a linear dependence in both $\log{\QQ}$ 
and $\log{\xinv}$. Furthermore, its extrapolation to smaller values in
$\log{\QQ}$ and $\log{\xinv}$, respectively, suggests the existence of
a common ``fixpoint'' (\xf,\QF) (cf. figs.~\ref{fig:q2fix},~\ref{fig:xfix}).
All this can be summarized in an ansatz for \FF\, which is linear in the 
double-logarithmic scaling variable $\xi$,
\begin{eqnarray}
  \FF(\xB,\QQ) &=& a + m\ \xi\ ,   \nonumber \\
  \xi          &=& \log{\frac{\QQ}{\QF}}\log{\frac{\xf}{\xB}}\ . 
  \label{eq:hypo}
\end{eqnarray}

\begin{figure}[hbtp]\centering
   \mbox{\epsfig{file=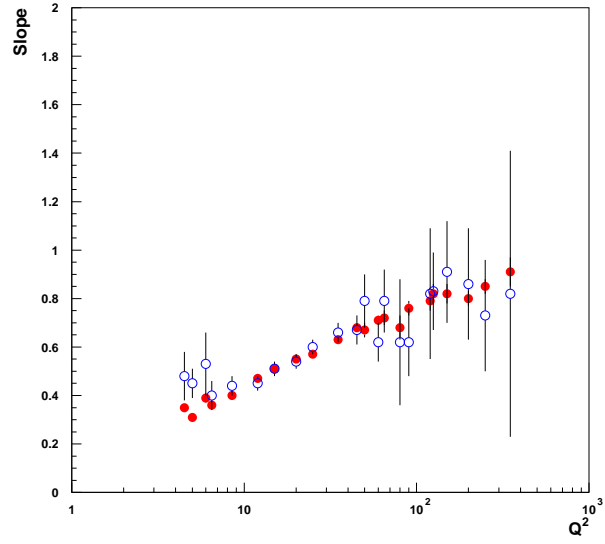,width=8.6cm}}
   \caption{\sl The open circles represent the directly measured slopes
            $u_1$, the full circles are the induced slopes for $\xf=0.074$
            and $a=0.078$. }
   \label{fig:slope}
\end{figure}

\begin{figure}[hbtp]\centering
   \mbox{\epsfig{file=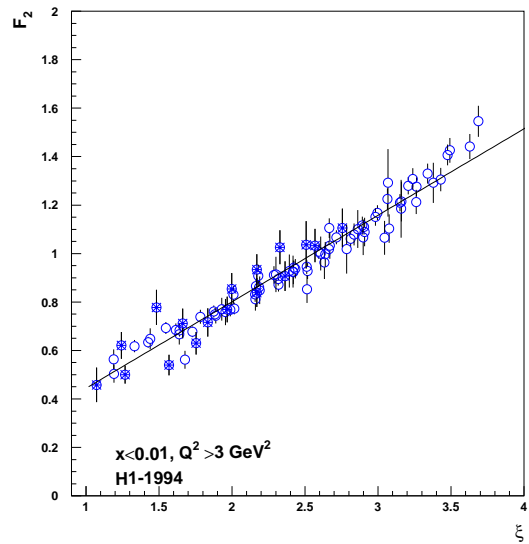,width=8.6cm}}
   \caption{\sl The \FF-data of H1-1994 is plotted versus
            the scaling variable $\xi$. The open circles represent the
            data for \QQ above 5 \GeV$^2$; the crossed circles correspond
            to the data with \QQ between 3 and 5 \GeV$^2$.}
   \label{fig:f2xi}
\end{figure}

This simple form, when confronted with the data given in (\ref{eq:data}),
implies 
\begin{eqnarray}
    a + m\log{\frac{\QQ}{\QF}}\log{\frac{\xf}{v(Q^2)}}
                                        &=& u_0(\QQ)\ , \nonumber \\
                 m\log{\frac{\QQ}{\QF}} &=& u_1(\QQ)\ .
    \label{eq:test}
\end{eqnarray}
The first of these equations can be cast into the form 
\begin{equation}
  \frac{u_0-a}{ \log{(\xf/v)} } = u_1\ , \nonumber
\end{equation}
thus allowing the comparison between the directly measured slope
$u_1$ and the one obtained from the point ($v,u_0$)
and the fixpoint (\xf,$a$). This is illustrated in fig.~\ref{fig:slope}.
The two parameter pairs ($a$,\xf) and ($m$,\QF) are correlated, as is obvious
from eq.~(\ref{eq:hypo}). They are chosen as follows :
\beq
   a = 0.078\ ,\ m = 0.364 \ ,\ \xf = 0.074\ ,\ \QF = 0.5\ \mbox{GeV}^2\ .
   \label{eq:para}
\eeq
Using the H1-data \cite{h12} with uncorrelated errors the measured structure
function \FF\ is plotted in fig.~\ref{fig:f2xi} as a function of the scaling 
variable $\xi$ computed from \xB and \QQ\ with the parameter values for \xf\ and
\QF\ as given in eq.~(\ref{eq:para}). The $\chi^2/dof$ is 83/72.
It turned out that even the data below 5 \GeV$^2$ are well described by 
eq.~(\ref{eq:hypo}). We have checked that also the other data on \FF\ 
(c.f. \cite{zeus,h11}) are well described by the same set of parameters.
%
\section{Constraints from unitarity}
%
No rigorous bound is known on the asymptotic behaviour of the structure
function $F_2(x,Q^2)$ as $x \rightarrow 0$, with $Q^2$ fixed. 
However, within the 
framework of the parton model constraints on the allowed growth of
parton densities at small $x$ can be obtained by considering the 
Froissart bound \cite{fross} on the total proton-proton cross section 
at large energies,   
\beq\label{frma}
\sigma_{tot}(s) \leq \frac{\pi}{m_{\pi}^2}\left(\ln{{s\over s_0}}\right)^2\ .
\eeq
Here $m_{\pi}$ is the pion mass, $s$ is the center-of-mass energy squared
and $s_0$ is an unknown constant. The total cross section 
measured at the Tevatron is of order $\pi/m_{\pi}^2 \sim 100$ mb. 

The cross section for the production of particles with high tranverse
momentum, $p_{\perp}^2 > \mu^2 \gg \Lambda_{QCD}^2$, 
can be calculated perturbatively.
Clearly, this part of the cross section, $\sigma_{pert}$, must also
satisfy the Froissart bound, i.e.
\beq\label{dtbound}
\sigma_{pert}(s, p_{\perp}^2 > \mu^2) < \sigma_{tot}(s) \ .
\eeq
Here $\sigma_{pert}$ is evaluated in terms of parton cross
sections and parton densities. Eq. (\ref{dtbound}) then yields a consistency
condition for the behaviour of the parton densities at small $x$.

In the parton model the dominant contribution to $\sigma_{pert}$ is
elastic gluon-gluon scattering.
Consider the production of two gluons with transverse momentum
$p_{\perp}\pm \Delta p_{\perp}$ and rapidities $y_3\pm \Delta y$ and
$y_4 \pm \Delta y$, respectively. Other partons in the final state
are summed over.  
The corresponding cross section reads in lowest order
perturbation theory (cf. \cite{stir})
\beq\label{d2jet}
\Delta\sigma_{gg} \simeq C \alpha_s(\mu)^2
x_1 g(x_1, \mu) x_2 g(x_2,\mu) {1\over p_{\perp}^4}
F(\theta^*) \Delta p_{\perp}^2 \Delta y^2 \ .
\eeq
Here $x_1$ and $x_2$ are the momentum fractions of the gluons in the initial
state, and $g(x,\mu)$ is the gluon density at scale $\mu$; $C$ is a constant
and $F$ is the averaged squared matrix element of the gluon-gluon
cross section, which depends only on $\theta^*$, the scattering angle
in the gluon-gluon center-of-mass system. Note, that the multiplicity
factor, which connects inclusive and exclusive cross sections,
is $(1 + \cal{O}(\alpha_s))$ in eq.~(\ref{d2jet}).
In the special case $y_3+y_4 =0$, one has
\beq\label{xs}
x_1 = x_2 \equiv x = {2 p_T\over \sqrt{s}}\ .
\eeq
The cross section (\ref{d2jet}) is a tiny fraction of the total cross
section, $\Delta \sigma_{gg} < \alpha_s^2/\mu^2 \sim 1 \mu$b.  
Its dependence on the center-of-mass energy $s$ scales like
$(x g(x,\mu))^2$, with $x=2p_{\perp}/\sqrt{s}$. Clearly, if the gluon
density satisfies the bound
\beq\label{gbound} 
x g(x,\mu) < B \ln{{1\over x}}\ ,
\eeq
where $B$ is a constant, then the corresponding bound on
$\Delta \sigma_{gg}$ and the Froissart bound scale with $s$ in the
same way, i.e. like $(\ln{s})^2$. Hence, $\Delta \sigma_{gg}$ will
always remain far below the Froissart bound. 

One may hope to derive a stronger bound on the gluon density based
on a complete evaluation of the perturbative cross section $\sigma_{pert}$.
Already the integration over the rapidities of the two final state gluons
yields another factor of $(\ln{s})^2$. Contributions with additional
gluons in the final state yield further powers of $\ln{s}$ which
eventually build up the full BFKL ladder \cite{lip}. It is conceivable
that these perturbative corrections can be absorbed in a properly
defined gluon density for which the bound (\ref{gbound}) may then apply.
A similar analysis could be carried out for deep-inelastic scattering
where one has only one gluon in the initial state. In this way it may
be possible to obtain a true bound on the structure function $F_2$ at
small $x$. This, however, is beyond the scope of our paper. In any case,
it is clear that with a logarithmic growth of the gluon density the 
perturbative cross section can be extrapolated several orders of
magnitude beyond present center-of-mass energies before it possibly
reaches the Froissart bound.

One can readily evaluate the contribution of photon-gluon fusion to the
structure function $F_2$ for a gluon density saturating the bound
(\ref{gbound}).  One finds (cf. \cite{field}),
\beq\label{fusion}
\Delta F_2(x,Q^2) = A + {\alpha_s\over 3\pi}\sum_q e_q^2 B 
\ln{{Q^2\over Q_0^2}} \ln{{x_0\over x}} ,
\eeq
where the sum extends over all quarks with masses small compared to $Q$,
and the constant $A$ depends on the renormalization scheme. 
Eq.~(\ref{fusion}) is identical to the fit (\ref{eq:hypo}) of $F_2$,
obtained by the phenomenological analysis in sect.~2, 
if parameters are properly identified.
>From the slope $m=0.364$ one obtains $B\simeq 3$.
At small $x$ the gluon density is large, and the structure function is 
likely to be dominated by photon-gluon 
fusion.\footnote{A similar fit to $F_2$, with $\ln{1/x}$ replaced by a small
power $x^{-\lambda}$ has been performed in \cite{heb}.} 

Eq. (\ref{fusion}) relates the behaviour of $F_2$ at small $x$ to a
constraint on the gluon density obtained from the total proton-proton 
cross section.  Our discussion has been based on a comparison of cross sections 
in perturbation theory to leading order, and the effect of higher
order corrections is not clear.
It is conceivable that, at fixed $Q^2$, $F_2(x,Q^2)$ continues to rise
like $\ln{1\over x}$ down to asymptotically small values of $x$. 
In fact, such a
behaviour has been predicted by Bjorken. His starting point is 
an expression for the photon wave function renormalization constant 
derived by Gribov \cite{grib},
\beq
1-Z_3 = {\alpha\over 3\pi}\int ds \frac{s R(s)}{(Q^2+s)^2}\ ,
\eeq
where $R(s)$ is the hadronic cross section in $e^+e^-$-annihilation
normalised to the $\mu$-pair cross section. Based on the
aligned-jet picture of deep inelastic scattering he then obtains for
the inclusive structure function \cite{bjo}
\beq\label{bjb}
F_2(x,Q^2)\ \propto\ \bar{R}\ \ln{1\over x}\ .
\eeq
This corresponds to a gluon density saturating the bound (\ref{gbound}).
At present, however, we do not know whether eqs.~(\ref{bjb}) or (\ref{fusion})
represent the correct behaviour of $F_2(x,Q^2)$ at
asymptotically small values of $x$. There is neither a proof that such
a behaviour will ever be reached nor can it be excluded that it might
set in already at moderate values of $x$.
%
\section{Comparison with double-asymptotic scaling}
%

So far we have discussed the Regge limit, i.e.  $x \rightarrow 0$
with $Q^2$ fixed. In this limit the double-logarithmic
scaling form (\ref{eq:hypo}) 
of the structure function \FF\ may be correct. However, in an appropriate
simultaneous limit $Q^2 \rightarrow \infty$ and $x \rightarrow 0$
(see below), the double-logarithmic scaling form is expected to be incorrect.
In this limit, for appropriate boundary conditions, the structure 
function is given by an asymptotic solution
of the standard evolution equations \cite{ynd}, which was found more
than 20 years ago \cite{der}, and which has recently been thoroughly 
studied by Ball and Forte \cite{bf1,bf2}. This solution is known to 
correspond to a summation of all terms of the form
\beq\label{series}
\left(\alpha_s \ln{{Q^2\over \Lambda^2}} \ln{{1\over x}} \right)^n\ .
\eeq
Hence, at small $x$, it increases faster than any power of $\ln{1\over x}$. The 
double-logarithmic form (\ref{eq:hypo}) corresponds to the first term in this 
series. A comparison of these two expressions with data on $F_2$ 
therefore tests for the evidence of terms more singular than $\ln{1\over x}$
at small $x$.

\begin{figure}[hbtp]\centering
   \mbox{\epsfig{file=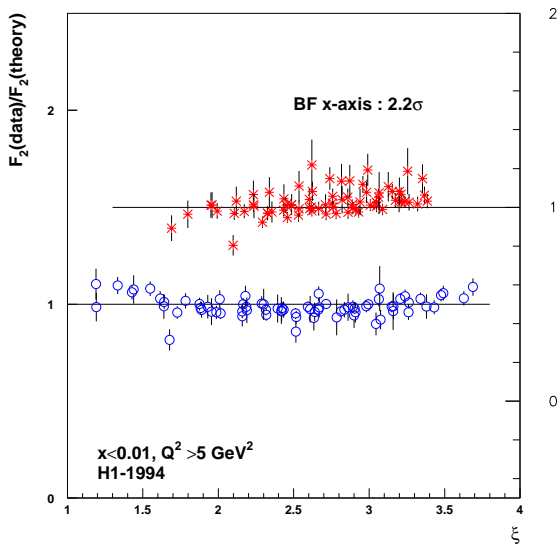,width=8.6cm}}
   \caption{\sl Comparison of \FF-data with the Ball-Forte fit (right
                vertical scale) and the double-logarithmic fit (left
                vertical scale); the horizontal axis corresponds to
                2.2 $\sigma$ and $\xi$, respectively.}
   \label{fig:dhbf}
\end{figure} 

The asymptotic solution is conveniently expressed in terms of the
variables \cite{bf2}
\beq
\sigma = \left(\ln{{x_0\over x}}\ln\left({\alpha_s(Q_0)\over
\alpha_s(Q)}\right)\right)^{1/2}\quad,\quad
\rho = \left(\ln{{x_0\over x}}\right/\left.\ln{\left({\alpha_s(Q_0)\over
\alpha_s(Q)}\right)}\right)^{1/2}\ ,
\eeq
where $\alpha_s$ is the two-loop QCD coupling,
\beq
\alpha_s(Q)={4\pi\over \beta_0\ln{Q^2\over \Lambda^2}}\left(1 -
{\beta_1\over \beta_0^2}{\ln\ln{Q^2\over \Lambda^2}\over 
\ln{Q^2\over \Lambda^2}}\right)\ .
\eeq
The first two coefficients of the $\beta$-function and other relevant
parameters read \cite{bf2}
\begin{eqnarray}
\beta_0 = 11 - {2\over 3}n_f\ ,\quad
\beta_1 = 102 - {38\over 3}n_f\ ,\quad
\gamma^2 = {12\over \beta_0}\ ,\quad
\delta = {1\over \beta_0}\left(11 + {2\over 27}n_f \right)\ ,\nn \\
\epsilon_+ = {1\over \beta_0}
    \left(3{\beta_1\over \beta_0}+{103\over 27}n_f\right)\ ,\quad
\epsilon_- = {78\over \beta_0\gamma^2}\ .\qquad\qquad 
\end{eqnarray}
At large values of $\sigma$ the structure function $F_2$ can now be written
as \cite{bf2}
\begin{eqnarray}
F_2(\xB,\QQ) &=& C\ \exp{2\gamma\sigma}\  
                    \exp{\left(-\delta {\sigma\over \rho}
         - {1\over 2} \ln{\gamma\sigma} - \ln{\rho\over \gamma}\right)} 
         \nonumber \\
 &&\ \times\left(1 + \left(\epsilon_+ + \epsilon_-\right)\alpha_s(Q)
     -\epsilon_+ \alpha_s(Q_0)\right){\rho\over \gamma} 
     \left(1+\cal{O}({1\over \sigma}) + \cal{O}({1\over \rho})\right)\ .
\label{f2bf}
\end{eqnarray}
Here $C$ is an unknown normalization constant. For large values of $\sigma$
and $\rho$ the first exponential gives the dominant contribution,
\beq
F_2(x,Q^2) \simeq C\ \exp{\left(2\gamma\sigma\right)}\ .
\eeq
In this limit the structure function only depends on $\sigma$, which is 
referred to as double-asymptotic scaling.

Eq. (\ref{f2bf}) gives $F_2$ in terms of four a priori unknown constants,
$\Lambda$, $Q_0$, $x_0$ and $C$. It is known that a proper choice of these
parameters yields a good description of the measured data on $F_2$ in the
range of small $x$ and large $Q^2$ \cite{bf2,h12}. Using the optimised 
values of \cite{h12}, the constancy of the ratio between the data and the
theoretical prediction (\ref{f2bf}) is clearly borne out in 
fig.~\ref{fig:dhbf} 
and quantified by $\chi^2/dof$ = 93/69. Eq. (\ref{f2bf}) has been used
for the number of flavours $n_f= 4$. A more sophisticated treatment 
incorporating \QQ-dependent threshold effects has not been attempted.

\begin{figure}[hbtp]\centering
   \mbox{\epsfig{file=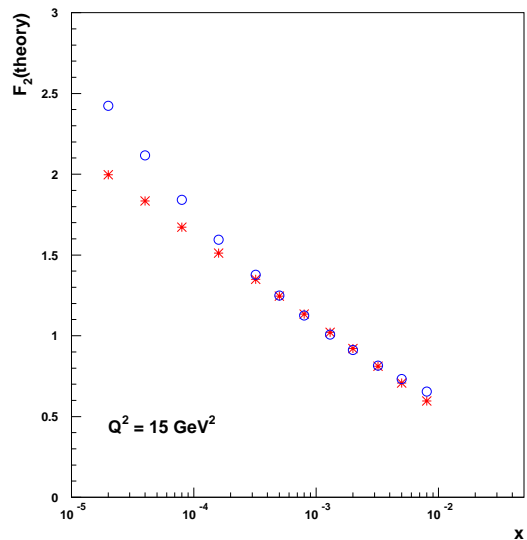,width=8.6cm}}
   \caption{\sl The structure function $F_2$ for $Q^2 = 15\ GeV^2$ for 
                the Ball-Forte fit (open circles) and the 
                double-logarithmic fit (stars) extrapolated to values
                of $x$ one decade below the HERA range. }
   \label{fig:extra}
\end{figure} 

What does this mean concerning the validity of double-asymptotic scaling?
As we have seen in sect. 2 (cf. also fig. \ref{fig:dhbf}), the data are 
equally well described by just the first term in the sum of double logarithms 
given in eq. (\ref{series}). Hence, the characteristic feature of
double-asymptotic scaling, a growth stronger than any power of
$\ln{1\over x}$ cannot be inferred from the present HERA data. 
This more singular behaviour should become visible if, at given
$Q^2$, the range in $x$ is extended by one order of magnitude. 
As fig.~\ref{fig:extra} illustrates, a clear difference is then expected 
between the double-logarithmic fit (\ref{eq:hypo}) and the 
double-asymptotic form (\ref{f2bf}).
%
\section{Conclusions}
%
We have shown that data on the structure function $F_2(x,Q^2)$ at small $x$ 
and large $Q^2$ published up to date are consistent with a
logarithmic growth in ${1\over x}$ as well as $Q^2$. This is of interest
for several reasons.

No rigorous bounds are known on the asymptotic behaviour of $F_2(x,Q^2)$
in the Regge limit, i.e. as $x\rightarrow 0$ with $Q^2$ fixed.
However, according to our discussion in sect.~3,
a logarithmic growth of $F_2$ can be extrapolated to smaller values of $x$ by
many orders of magnitude without getting into conflict with the
Froissart bound. Such a logarithmic increase with ${1\over x}$ may even be
the correct asymptotic behaviour of the structure function.

In the double-asymptotic regime of small $x$ and large $Q^2$ perturbative
QCD predicts double-asymptotic scaling for the structure function $F_2$,
given sufficiently soft input distributions. This implies a growth
stronger than any power of $\ln{1\over x}$. So far, however, the data are
still consistent with double-logarithmic scaling corresponding to an
increase like $\ln{1\over x}$. This may mean that at HERA the small-x
regime has not yet been reached, where the strong growth expected on
the basis of perturbative QCD will become visible. 

More precise measurements may differentiate between 
double-logarithmic scaling, double-asymptotic scaling and the even more
singular BFKL power behaviour. As discussed in sect. 4, these differences
should become manifest if, for given $Q^2$, the range in $x$ is extended
by one order of magnitude. This corresponds to an increase in the
center-of-mass energy squared by one order of magnitude, which could be 
reached at future colliders, such as
LEP$\otimes$LHC or at a 500 GeV Linear Collider$\otimes$HERA.

We would like to thank J. Bartels, P. V. Landshoff,  F. Schrempp 
and P. M. Zerwas for valuable discussions.

\newpage

%

%
%
\end{document}